\documentclass[12pt,twoside]{article}   

\usepackage{multirow}
\usepackage{amsmath}

\usepackage[super,sort,comma]{natbib}

\usepackage[section]{placeins}   %

\usepackage{graphicx}

\makeatletter \renewcommand\@biblabel[1]{$^{#1}$} \makeatother
\newcommand{\cen}[1]{\begin{center} #1 \end{center}}

\usepackage[mathlines]{lineno}

\usepackage[mathlines]{lineno}

\usepackage{hyperref}
\hypersetup{ colorlinks,
    citecolor=blue,
    filecolor=blue,
    linkcolor=blue,
    urlcolor=blue
}

\usepackage{xcolor}

\definecolor{gray}{rgb}{0.6,0.6,0.6}
\definecolor{red}{rgb}{0.85,0,0}
\definecolor{green}{rgb}{0,0.85,0}
\definecolor{blue}{rgb}{0,0,0.85}
\definecolor{beige}{rgb}{0.92,0.87,0.78}
\usepackage[all]{hypcap}    

\begin{document}


\cen{\sf {\Large {\bfseries Supersonic gas curtain based real-time ionization profile monitor for hadron therapy } \\  
\vspace*{10mm}
Narender Kumar$^{1 ,2}$, Milaan Patel$^{+,1 ,2}$, William Butcher$^{1 ,2}$, Hao Zhang$^{1 ,2}$, Oliver Stringer$^{1 ,2}$, Joseph Wolfenden$^{1 ,2}$, Carsten P Welsch$^{1 ,2}$} \\
$^1$Cockcroft Institute, Warrington, WA4 4AD, United Kingdom. \\
$^2$Physics Department, University of Liverpool L69 7ZE, United Kingdom
\vspace{5mm}\\
}

\pagenumbering{roman}
\setcounter{page}{1}
\pagestyle{plain}
$^+$Author to whom correspondence should be addressed. email: milaan.patel@liverpool.ac.uk \\

\begin{abstract}
\noindent {\bf Background:} Accurate control and monitoring of the beam is essential for precise dose delivery to tumor tissues during radiotherapy. Real-time monitoring of ion beam profiles and positions improves beam control, patient safety, and treatment reliability by providing immediate feedback. This becomes even more critical in FLASH therapy, where the short corrective window during high-dose delivery demands precise beam control. \\
{\bf Purpose:} Existing devices are often limited to in vitro calibration or focus on monitoring a single parameter during treatment. This study aims to develop a device that can simultaneously monitor beam position, profile, current, and energy in real-time, without perturbing the beam, using a supersonic gas curtain system. \\ 
{\bf Methods:} A supersonic gas curtain beam profile monitor was developed at the Cockcroft Institute to assess its performance and suitability for applications in hadron-beam therapy. The system was integrated with one of the beamlines of the Pelletron accelerator at the Dalton Cumbrian Facility, Whitehaven, United Kingdom and 2D transverse beam profile measurements of carbon beams were conducted. \\
{\bf Results:} The monitor successfully measured the beam profiles within 100 ms to 1 s across various beam currents (1 - 100 nA), energies (12 - 24 MeV), and charge states ($\mathrm{C^{2+}}$-$\mathrm{C^{5+}}$) of carbon ions. Recorded data was used to estimate detector performance by introducing a parameter called 'detection limit' to quantify sensitivity of the monitor, identifying the threshold number of ions required for detection onset. A method to quantify sensitivity under different beam conditions is then discussed in detail, illustrated with an example case of FLASH beam parameters.  \\
{\bf Conclusions:} This proof-of-concept (POC) study demonstrates the performance of the gas curtain-based ionization profile monitor for 2D transverse beam profile measurement of carbon ions. The device's sensitivity is quantified and evaluated against an example case for FLASH conditions. The paper also discusses options to enhance performance of the monitor. \\

\end{abstract}

\pagenumbering{arabic}
\setcounter{page}{1}

\section{Introduction}

Radiotherapy has consistently aimed to improve dose delivery to tumors while sparing normal tissue. This requires conformal dose delivery to enhance Tumor Control Probability (TCP) and reduce Normal Tissue Complication Probability (NTCP) \cite{Baumann2005}. Ion beams offer localized dose delivery through the benefit of the Bragg peak \cite{IAEA2006}. Protons are preferred choice in clinical practice, along with heavier ions, such as carbon, for further enhancement of dose conformity and linear energy transfer (LET) \cite{Mohamad2017}. The Bragg peak characteristic is common to all ions. Since its first demonstration in the 1940s, hadron therapy has advanced significantly from research laboratories to clinical treatment \cite{Tsuboi2020}. The latest modality, FLASH therapy, shows promising results in sparing normal tissue with some reduction in tumor control \cite{Wilson2012}. Naturally, there is a strong push towards clinical implementation of FLASH hadron beams, with upcoming facilities dedicated towards delivering hadron beams under FLASH conditions \cite{LhARA2022}. Successful clinical implementation of hadron beam therapy requires seamless integration of precise beam control from the accelerator side and accurate dose measurement from the clinical side, where beam diagnostics play a crucial role. 

Beam diagnostics can be achieved either from the accelerator end - where beam current, energy, and profile are critical - or from the clinical end, where dose distribution provides comprehensive beam characterization. Diagnostics from the accelerator end provides the necessary feedback to control the beam and often serves as the only beam diagnostics during in-vivo treatments where feedback from the dosimetry can lack critical information about the dose distribution, beam profile and dose rates. There is a need for advanced diagnostics for FLASH beams, where the instantaneous dose rate is exceptionally high, and the entire treatment occurs within a few interactions of few seconds \cite{Jolly}. A real-time feedback can thus ensure precision in dose delivery by allowing for implementing corrective measures to prevent overdosing.

Conventional dosimetry techniques are insufficiently evolved to provide real-time feedback on the dose profile, non-invasively, for FLASH hadron beams. Ionization Chambers (IC), widely used, suffer from recombination issues \cite{Petersson2017,Patriarca2018,Jorge2019,Beyreuther2019}, while solid-state diodes can exhibit over-response \cite{Neira2019,Jursinic2013}. Radiographic films, also commonly used for FLASH dosimetry, offer high accuracy but can under-respond in specific conditions and lack real-time capabilities \cite{Kojima2003,Marti2010}. Active stimulated dosimeters provide real-time feedback and dose distribution but cannot be implemented for in-vivo treatments \cite{Benmakhlouf2017,Parwaie2018,Casar2019}. Detectors like Cherenkov monitors offer real-time feedback and full dose-depth information without disturbing the beam, yet they are primarily suited for electron/photon beams \cite{Axelsson_2011}. Graphite calorimeters, known for extreme accuracy in photon \cite{Seuntjens2009} and electron \cite{McEwen2009} beams, are being adapted as reference standards for ion beams \cite{Rossomme2013}, but they provide only total dose information without real-time capabilities. Given these limitations, there is ongoing effort to explore new techniques that can address some of these challenges. 

This paper discusses the progress on a Supersonic Gas Curtain-based real-time Ionization Profile Monitor (SGC-IPM) aiming to provide non-invasive, real-time transverse dose profile information for hadron beams. Measuring the beam profile is a step towards accurately monitoring the beam current distribution which can provide information about dose distribution. This instrument is a modified version of the Beam Gas Curtain (BGC) monitor originally developed for use on the High Luminosity Large Hadron Collider (HL-LHC) at CERN \cite{Salehilashkajani2022}. Unlike the BGC, which uses beam induced fluorescence, the SGC-IPM employs an approach that uses the ionizing interaction between neutral gas curtain and the particle beam using a unit called Ionization Profile Monitor (IPM) developed by our group to improve the sensitivity \cite{Tzoganis2014,Tzoganis2017}. The increased sensitivity achieved allows to capture real-time beam profile information as demonstrated by a comparative study using an electron beam \cite{Kumar2020}. The next step is to demonstrate the feasibility of beam profile measurements for hadron beams under conditions similar to medical accelerators, which is the primary objective of this study. The secondary objective is to quantify the detector sensitivity to co-relate the measurement with beam conditions for FLASH.

This work presents the experimental measurements of the transverse beam profile of carbon ions, measured by SGC-IPM within the therapeutic energy range. Measurements were conducted on one of the beamlines of the Pelletron accelerator at the Dalton Cumbrian Facility (DCF), UK, using carbon ions in charge states 2+, 3+, 4+, and 5+, with energies ranging from 12 to 24 MeV and currents varying from 2 to 100 nA. This paper describes the design features of SGC-IPM and the experimental methodology employed. Finally, the paper discusses current challenges and ongoing improvements planned to achieve the goal of real time dose profile monitoring.

\section{Methods}

\subsection{The supersonic gas curtain beam profile monitor}

\begin{center}
    \begin{figure}[h]
        \includegraphics[width=1\linewidth]{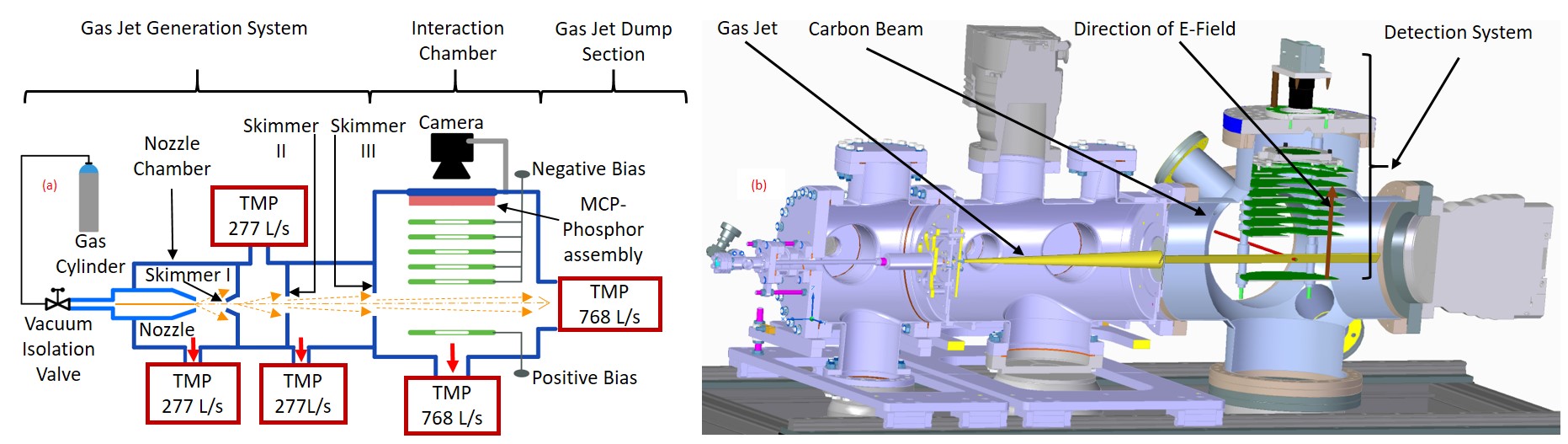}
        \caption{(a) Schematic and (b) 3d cross-sectional drawing of the supersonic gas jet monitor.}
        \label{fig:gas_jet_system}
    \end{figure}
\end{center}

SGC-BPM is a modular system comprising a supersonic gas curtain generation unit and a detection unit known as the Ionization Profile Monitor (IPM). Figure~\ref{fig:gas_jet_system}  (a) shows the schematics of the system from the perspective of looking along the beamline. The principle of supersonic curtain generation has been discussed previously \cite{Tzoganis2014,Tzoganis2017,Kumar2020}. A gas curtain is created through a series of steps, starting with the generation of a supersonic gas jet by adiabatic expansion of a high-pressure gas in to a vacuum. This is followed by extracting the supersonic core using skimmers and reshaping it into the gas curtain. The gas curtain is then injected into the chamber housing the IPM, perpendicular to the charged particle beam, with the surface inclined at 45$\mathrm{^o}$ with respect to the beam. The curtain acts as a screen of atoms/molecules. As the beam travels through the curtain, it ionizes the neutral atoms in the curtain in the shape of the beam transverse profile, with the spatial distribution of ions mirroring the intensity distribution of the beam. The ions are extracted from the curtain, maintaining their relative spatial distribution onto a Microchannel Plate (MCP) detector using a uniform electric field maintained by a set of extractor plates housed within the IPM, shown in green in figure \ref{fig:gas_jet_system} (color online). As the ions fall on the MCP, they produce a shower of secondary electrons, amplifying the signal by a factor of approximately $\mathrm{10^3}$. The electron shower is collected on a phosphor screen, which emits light that is imaged by a standard CMOS camera. 

In the above technique, the resolution of the beam profile generated is fundamentally limited to the minimum pore size of the MCP, which in our case is 25 µm in size. While the BGC monitor\cite{Salehilashkajani2022} has better resolution, IPM collects almost every ion that is created during ionization, thereby making is more sensitive. Moreover, the cross-section of ionization is generally much better as compared to the fluorescence cross-section. This increase in sensitivity significantly reduces the integration time required to collect sufficient charge to reconstruct the beam profile. Therefore, despite its complexity, the IPM is essential for achieving real-time beam monitoring and is used in the here-presented experiments. 

\subsection{Pre-experiment setup}

\begin{center}
\begin{table}[h]
    \centering
    \caption{Absolute pressures for different conditions for both Ar and N$_2$ jets}
    \label{tab:pressure}
    \begin{tabular}{p{1.2cm}p{1.2cm}|p{2.0cm}p{2.5cm}p{2.5cm}p{2cm}}
        \hline
        \multicolumn{2}{c}{Status} & \multicolumn{4}{c}{Pressure (mbar)} \\ 
        \hline
        Gas Jet & Beam & Nozzle Chamber & Skimmer Chamber (between 1st-2nd) & Skimmer Chamber (between 2nd-3rd) & Interaction Chamber \\
        \hline
        Off & Off & 3.98 $\times 10^{-8}$ & 7.10 $\times 10^{-9}$ & 3.00 $\times 10^{-9}$ & 1.73 $\times 10^{-8}$ \\
        On & Off & 4.84 $\times 10^{-3}$ & 5.60 $\times 10^{-6}$ & 4.60 $\times 10^{-7}$ & 2.13 $\times 10^{-8}$ \\
        On & On & 4.93 $\times 10^{-3}$ & 5.60 $\times 10^{-6}$ & 4.60 $\times 10^{-7}$ & 2.16 $\times 10^{-8}$ \\
        \hline
    \end{tabular}
\end{table}
\end{center}

Figure \ref{fig:gas_jet_system} (b) shows a 3D cross-sectional drawing of the SGC-IPM, composed of four chambers. Starting from the left, the nozzle chamber houses a 30 µm diameter nozzle operating at 5 bar gauge pressure to generate a highly underexpanded supersonic jet. The circular supersonic core is extracted into the second chamber through a 400 µm conical skimmer. It then passes through a 2 mm circular skimmer at the end of the second chamber, filtering the low-density halo and ensuring uniform cross-sectional density of the jet. The end of the third chamber has a slit skimmer (20 mm x 0.4 mm) mounted at a 45$\mathrm{^o}$ angle to generate an inclined gas curtain. The stream of gas is indicated in yellow (colour online) in figure \ref{fig:gas_jet_system} (b). After the interaction chamber, the gas jet reaches the dump section, a turbo-molecular pump, which efficiently removes the jet, minimizing the impact on the background vacuum level. Curtains of two different gases, nitrogen ($\mathrm{N_2}$) and argon (Ar), were generated for the measurements. Table \ref{tab:pressure} provides the vacuum levels in different chambers during operation with both, argon and nitrogen. 

\begin{figure*}
    \centering
    \includegraphics[width=1\linewidth]{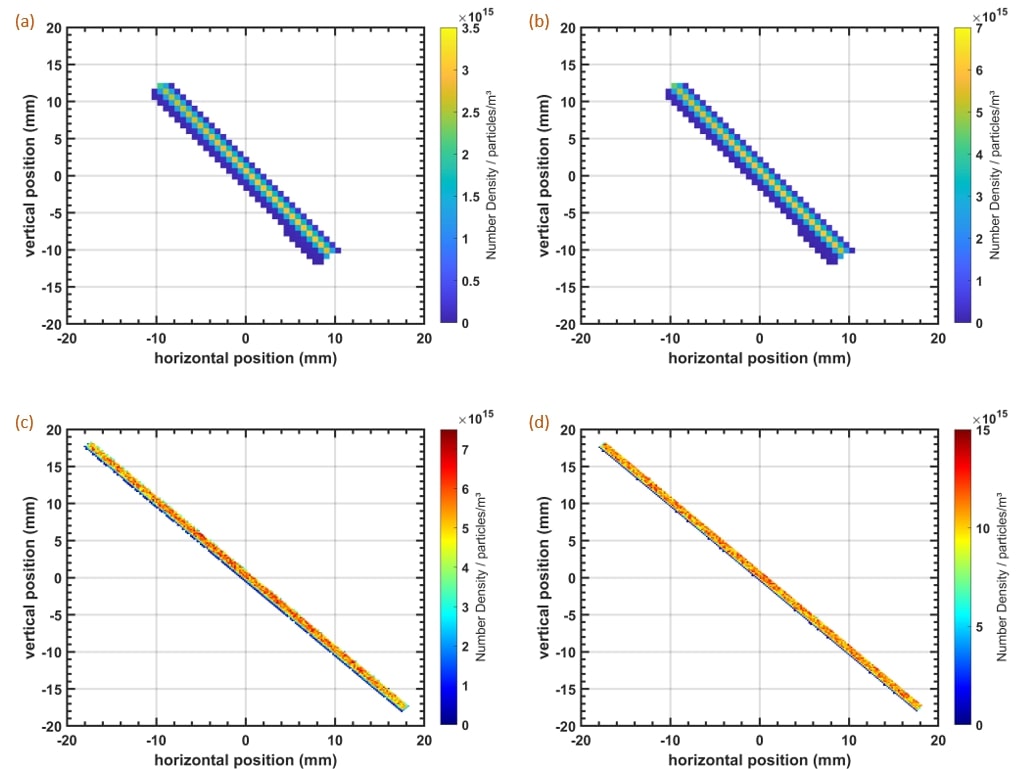}
    \caption{Supersonic gas jet density profiles at the interaction point: (a) \& (b) experimentally measured using moving gauge and (c) \& (d) simulated data, for nitrogen and argon gases respectively. \cite{Zhang2023} }
    \label{fig:gas_jet_density}
\vspace{10mm}
\end{figure*}

Figure \ref{fig:gas_jet_density} illustrates both the experimentally measured and calculated gas jet density profiles at the interaction point for nitrogen and argon gases using the scanning gauge technique \cite{Zhang2023}. The average gas density at the interaction point, with an injection pressure of 5~bar, is approximately $\mathrm{\approx 2.4 \times 10^{15}}$ molecules/m$\mathrm{^3}$ for nitrogen  and $\mathrm{\approx 5 \times 10^{15}}$ molecules/m$\mathrm{^3}$ for argon gas. The difference between experimental and simulated densities are due to the inconsistencies in accounting for skimmer interference and some unavoidable inaccuracies in measurement. For example, the absolute accuracy of a typical ion gauge used in scanning gauge technique is approximately 30~\%. The range of the scanning gauge used was limited to 10 mm in either direction which is the reason behind the small size of the curtain in figure~\ref{fig:gas_jet_density}.

The extractor plates of the IPM are negatively biased with voltages progressively decreasing from 200 V at the repeller (bottom) to -1,300 V at the MCP, generating a uniform extraction field of approximately 10 kV/m.  The MCP operates in ion detection mode with the rear end connected to the ground, and the phosphor screen (P47) biased at 3,000 V. Emission, peaking at 400 nm, from the phosphor screen is captured using a Basler ace CMOS camera (Model acA1920-40gm). The detection time is determined by the ion travel time to the MCP detector, which is simulated to be approximately 20 µs. The minimum integration time that can be achieved depends on the total number of ionization events, which varies with the current density of the beam and ionization cross-section of gas jet molecules. Current density is unique to a particular beam size and current while ionization cross-section varies with the energy. As a results the number of ionization, hence, the resulting signal strength varies with the beam energy and current as seen in our results shown later.

\subsection{Experiments}

\begin{figure} [h]
    \centering
    \includegraphics[width=0.8\linewidth]{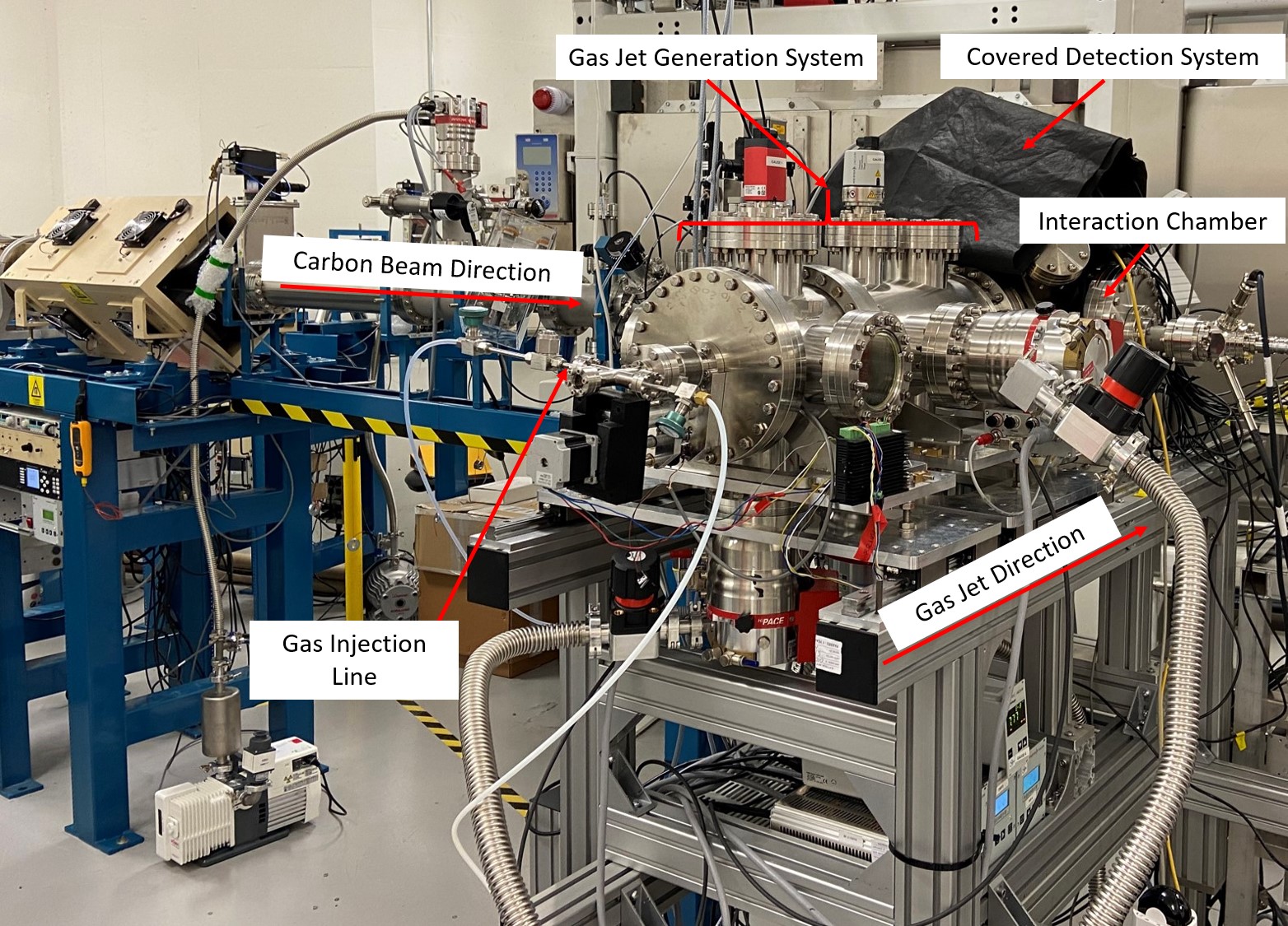}
    \caption{Image of SGC-IPM installed on the beam line at DCF.}
    \label{fig:integrated_setup}
\vspace{10mm}
\end{figure}

Figure \ref{fig:integrated_setup} shows the supersonic gas curtain monitor installed on the beamline of the pelletron accelerator at DCF. The interaction chamber is connected to the vacuum side of the accelerator. The ion beam enters the interaction chamber, passes through the IPM, and is then dumped onto a Tantalum foil (99.999\% pure) mounted at the end of the chamber. The foil acts as a beam dump with minimal residual activity. It was electrically isolated and connected to an ammeter, allowing it to be used as a beam current monitoring device. However, secondary electron loss from the foil could not be avoided due to the lack of a suppressor, so the foil was used only to monitor changes in the beam current during profile measurements. The absolute beam current was estimated using a Faraday cup installed upstream the beamline. 

Table \ref{tab:beam_current} summarizes the beam parameters for the carbon beam profile measurements. The beam was extracted from the DCF pelletron accelerator by maintaining the terminal voltage at approximately 4 MV. Beam energy was adjusted by selecting different charge states of the carbon ions by changing the magnetic field of the bending dipole magnet used for extraction. For a particular charge state $n$, the beam energy is determined by $E(MeV)= (n+1)V_T + V_s$, where $V_T$ is the terminal voltage and $V_s$ is beam extraction voltage of the ion source in MV. Thus, the beam energy $E$ varies in multiples of 4. For each beam energy, two sets of 2D transverse profile measurements were performed - one for each gas species by varying the beam sizes and currents. Due to operational limitations of the accelerator, stable beams could only be generated at specific currents for different charge states. Hence, beam profile were measured only for these discrete values. 

\begin{center}
\begin{table}[h]
    \centering
    \caption{Carbon beam parameters for beam profile measurements}
    \label{tab:beam_current}
    \begin{tabular}{cccc}
        \hline
        \multirow{2}{*}{Beam Energy E (MeV)} & \multirow{2}{*}{Charge State (q)} & \multicolumn{2}{c}{Beam Current (nA)} \\
        \cline{3-4}
        & & Argon Curtain & Nitrogen Curtain \\
        \hline
        12 & 2+ & 2.35 , 11.31 , 103.75 		& NA \\
        16 & 3+ & 10.5 , 11.31 , 99 , 103.75 & 2.00 , 93.00 \\
        20 & 4+ & 2.35 , 10.5 , 11.31 , 99 	& 2.00 , 19.95 \\
        24 & 5+ & 103.75 					& NA \\
        \hline
    \end{tabular}
\end{table}
\end{center}

\section{Results}

\subsection{Measurements and Data Analysis}

\begin{figure}
    \centering
    \includegraphics[width=0.45\linewidth]{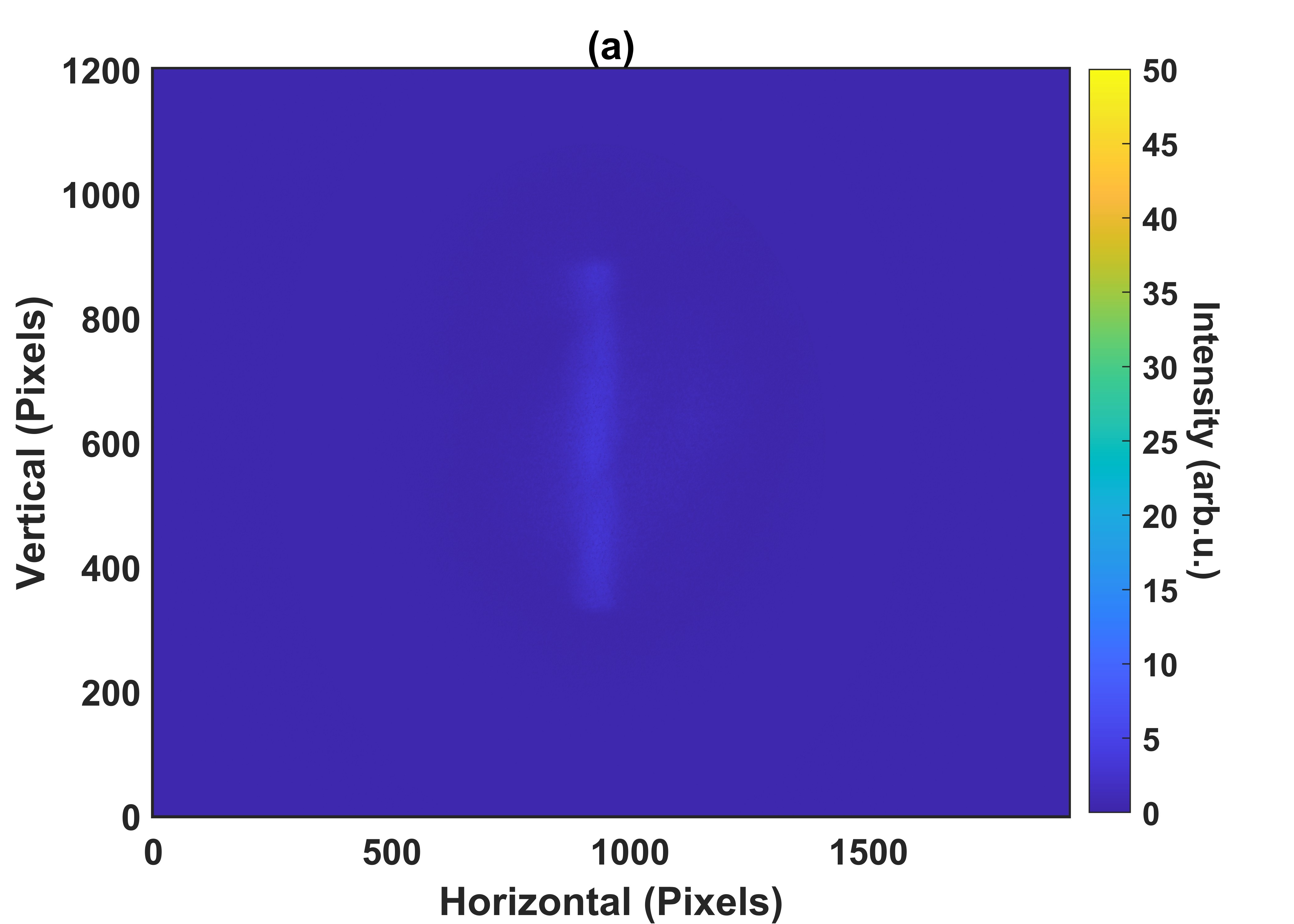}
    \includegraphics[width=0.45\linewidth]{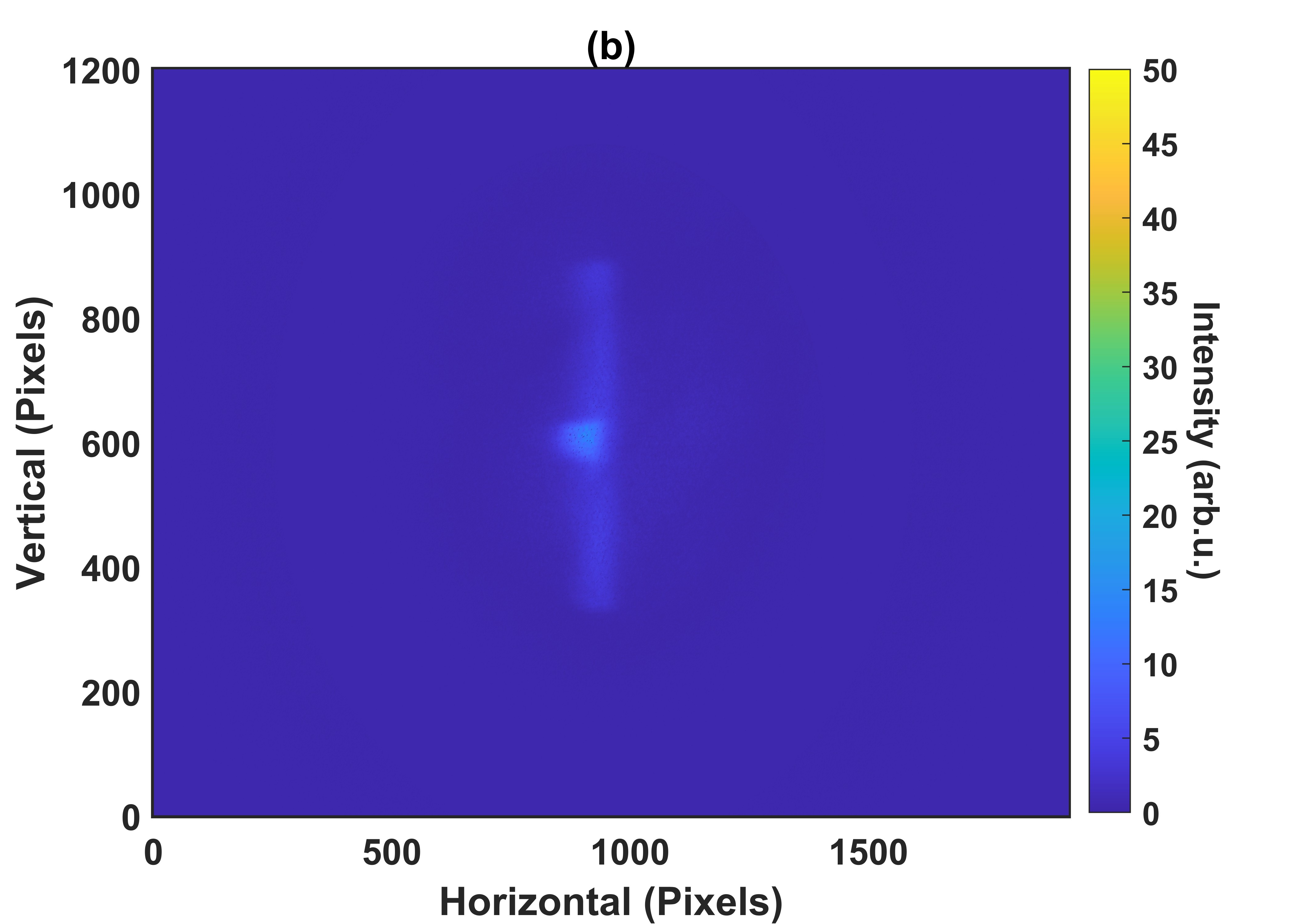}
    \caption{Images captured by IPM for 24 MeV C5+ beam with 100 ms integration time: (a) without gas curtain and (b) with gas curtain.}
    \label{fig:C5_100_ms_Ar}
\vspace{10mm}
\end{figure}

For beam parameters shown in Table \ref{tab:beam_current}, measurements were recorded with and without the gas curtain. Images for the 24 MeV, 100 nA carbon~5+ beam under both conditions are shown in figure~\ref{fig:C5_100_ms_Ar}. In the absence of gas curtain, a weak signal is observed due to the interaction of the carbon beam with the residual gas present in the interaction chamber. This appears as a vertical line along the beam path in figure \ref{fig:C5_100_ms_Ar}(a), with the intensity representing the integrated value of the beam intensity along the line of sight. In the presence of the gas curtain, a stronger signal showing the beam profile is observed at the location of the interaction between the gas jet and the carbon beam, as shown in figure \ref{fig:C5_100_ms_Ar}(b). Note that the beam profile is shifted from the original beam path. This shift occurs because the atoms/molecules in the curtain have directional velocity compared to the background gas, which is conserved upon their ionization and subsequent detection. To measure the beam intensity and profile, the contribution of the residual gas is subtracted, and a region of interest is defined around the beam.

\begin{figure*}
    \centering
    \includegraphics[width=0.45\linewidth]{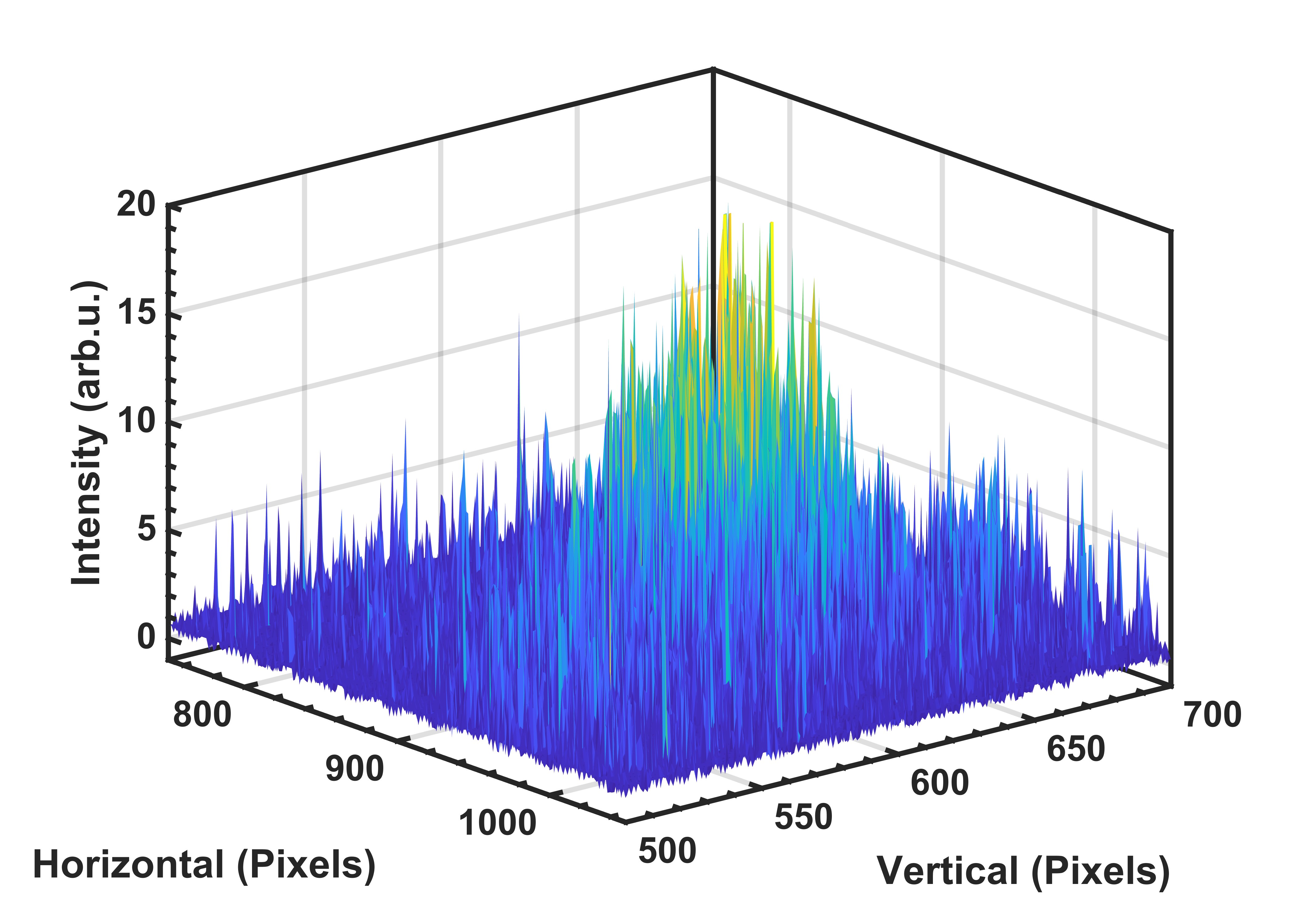}
    \includegraphics[width=0.45\linewidth]{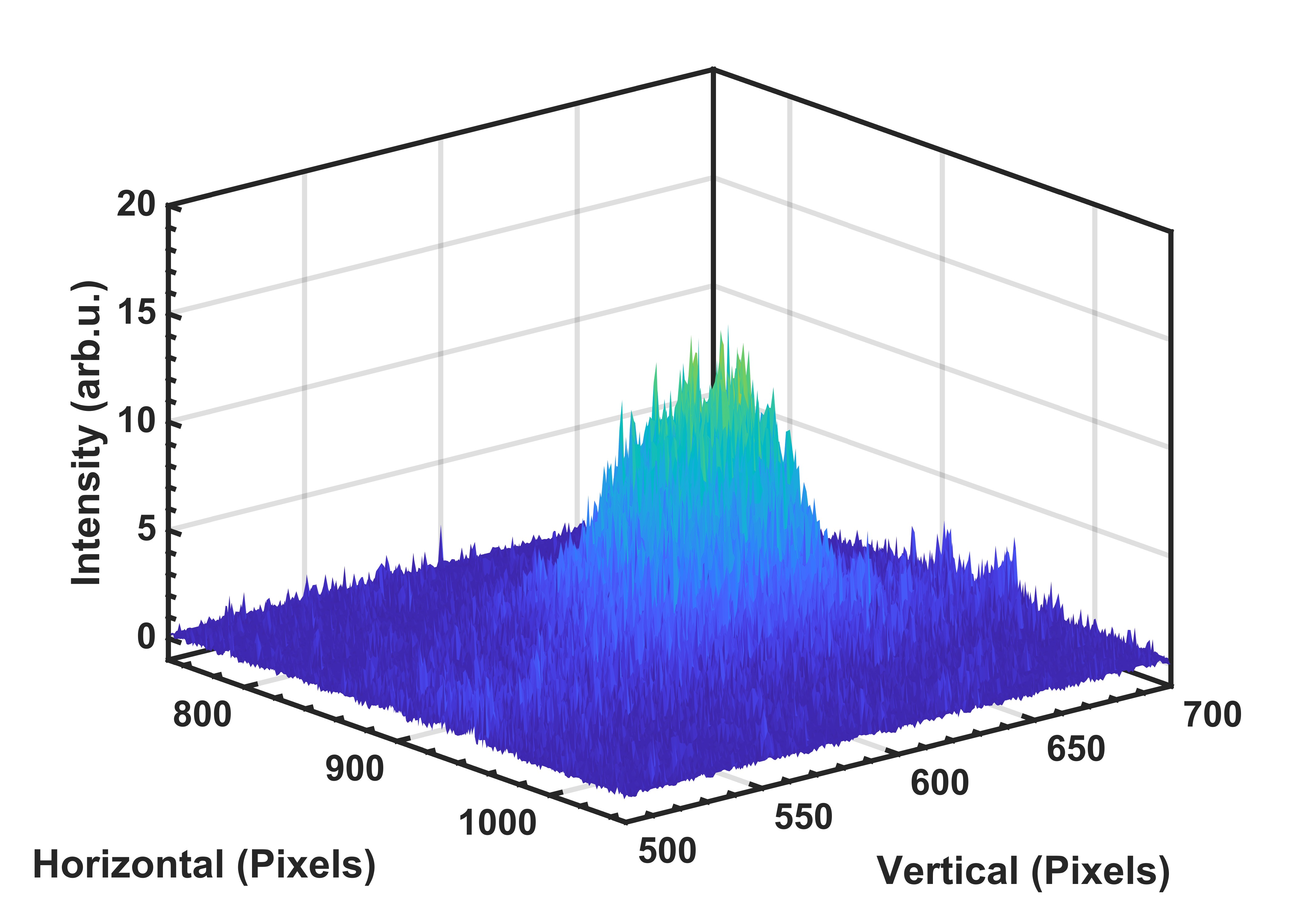}
    \includegraphics[width=0.45\linewidth]{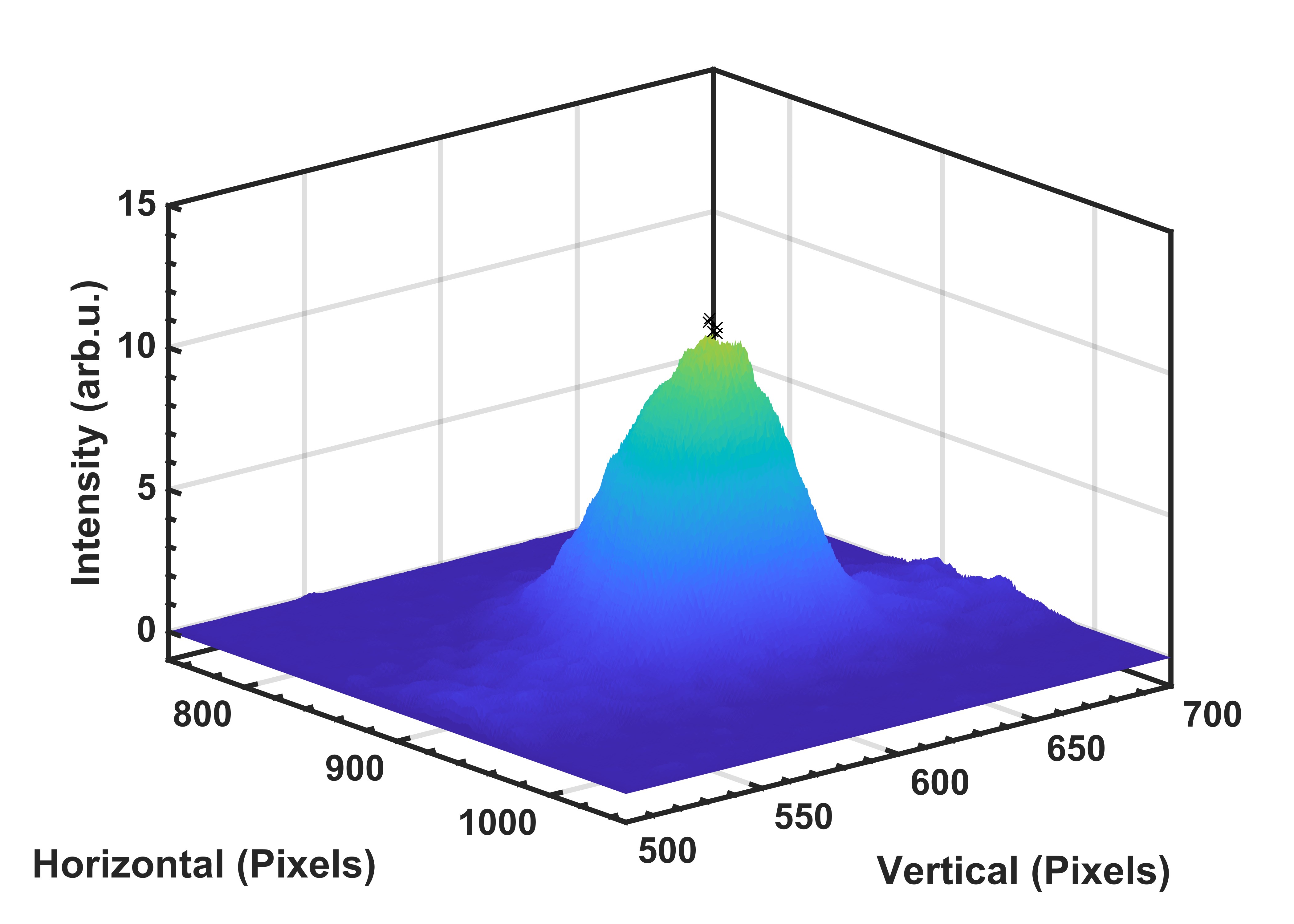}
    \includegraphics[width=0.45\linewidth]{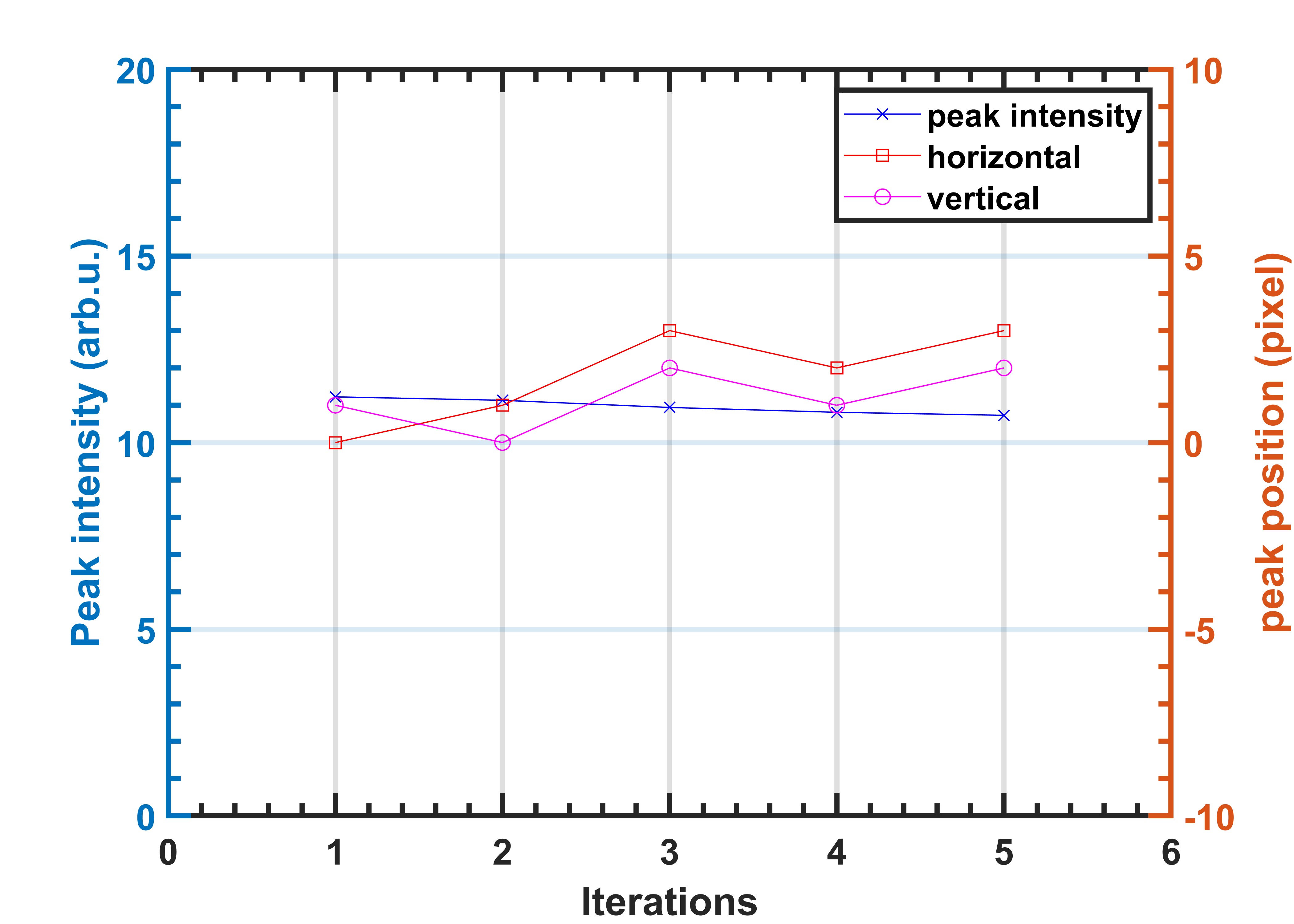}
    \caption{Beam profile of a 24 MeV C5+, $\sim$100 nA beam: (a) using a single image with 100 ms integration time, (b) averaging 10 images with 100 ms integration time, (c) after 5 iterations of smoothing, and (d) the variation in beam position and peak intensity during the smoothing process.}
    \label{fig:beam_profile}
\vspace{10mm}
\end{figure*}

Figure \ref{fig:beam_profile}(a) shows the spatial shape of the beam profile, generated using a single image of 100 ms integration time for a 24 MeV, $\sim$100 nA beam. The detector's dark noise averages around 5 counts, so a signal must exceed this level to be distinguishable from noise. This threshold limits the detection limit of the IPM, which will be discussed later.

Multiple images are averaged to reduce noise to 1 count. The number of images averaged depends on the dark noise of the detector and the desired confidence level in the measurement. The former varies with the camera sensor, while the latter is determined by clinical requirements. In this study, data analysis was conducted by averaging at least 10 images for each beam condition listed in Table \ref{tab:beam_current}. The beam profile using the average of 10 images is shown in figure \ref{fig:beam_profile} (b). 

The image is further processed by multipoint median smoothing over 5 iterations to remove data ripples. In this approach, pixels with noise (sharp peaks) are identified and replaced with the median of surrounding pixels in individual iterations. Parameters like peak height and peak position are monitored closely with each iteration to ensure that smoothing does not affect the profile. The resulting beam profile is shown in figure \ref{fig:beam_profile} (c) and the variation of beam parameters with each iterations is shown in figure \ref{fig:beam_profile} (d). 

\begin{figure}
    \centering
    \includegraphics[width=0.45\linewidth]{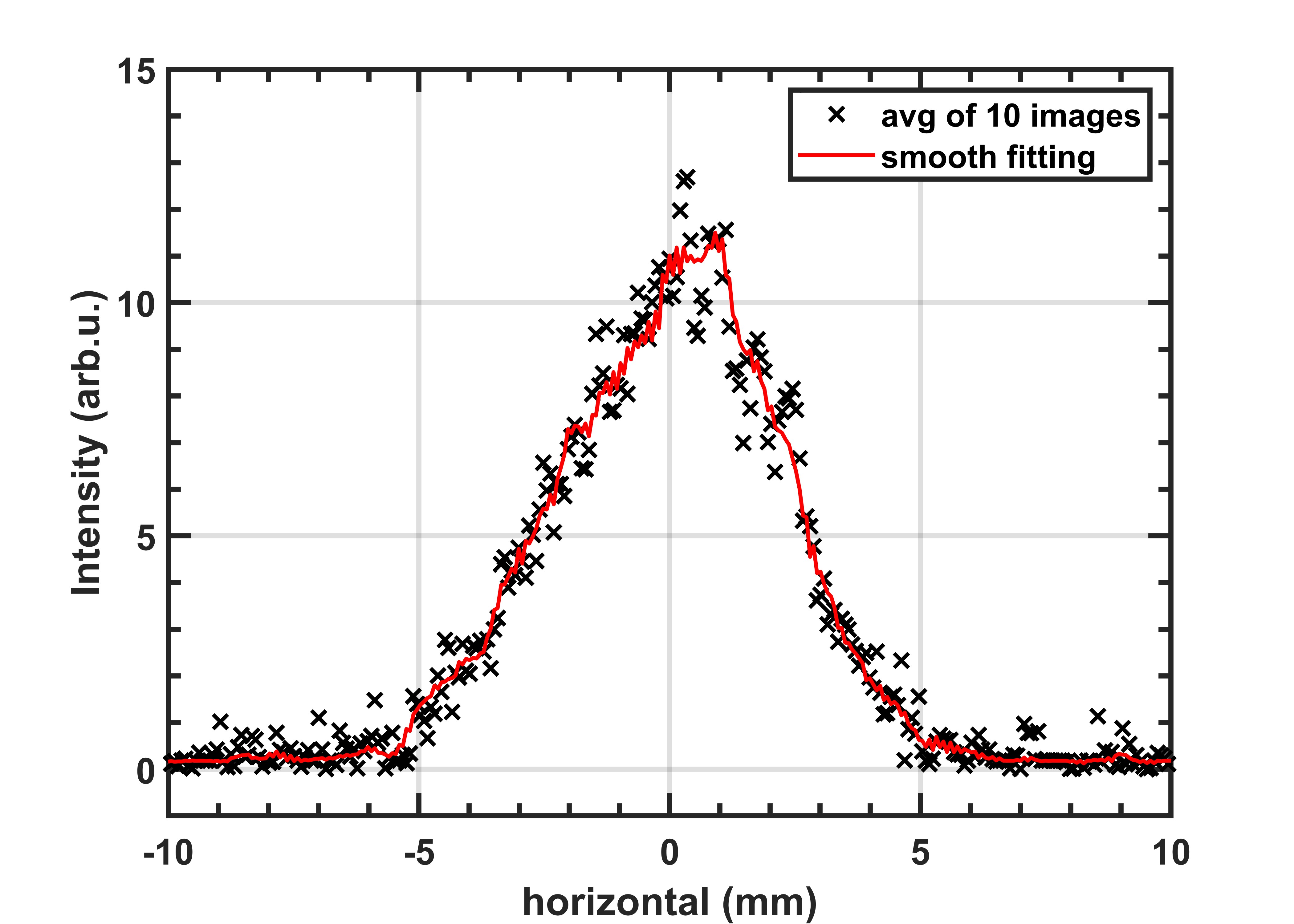}
    \includegraphics[width=0.45\linewidth]{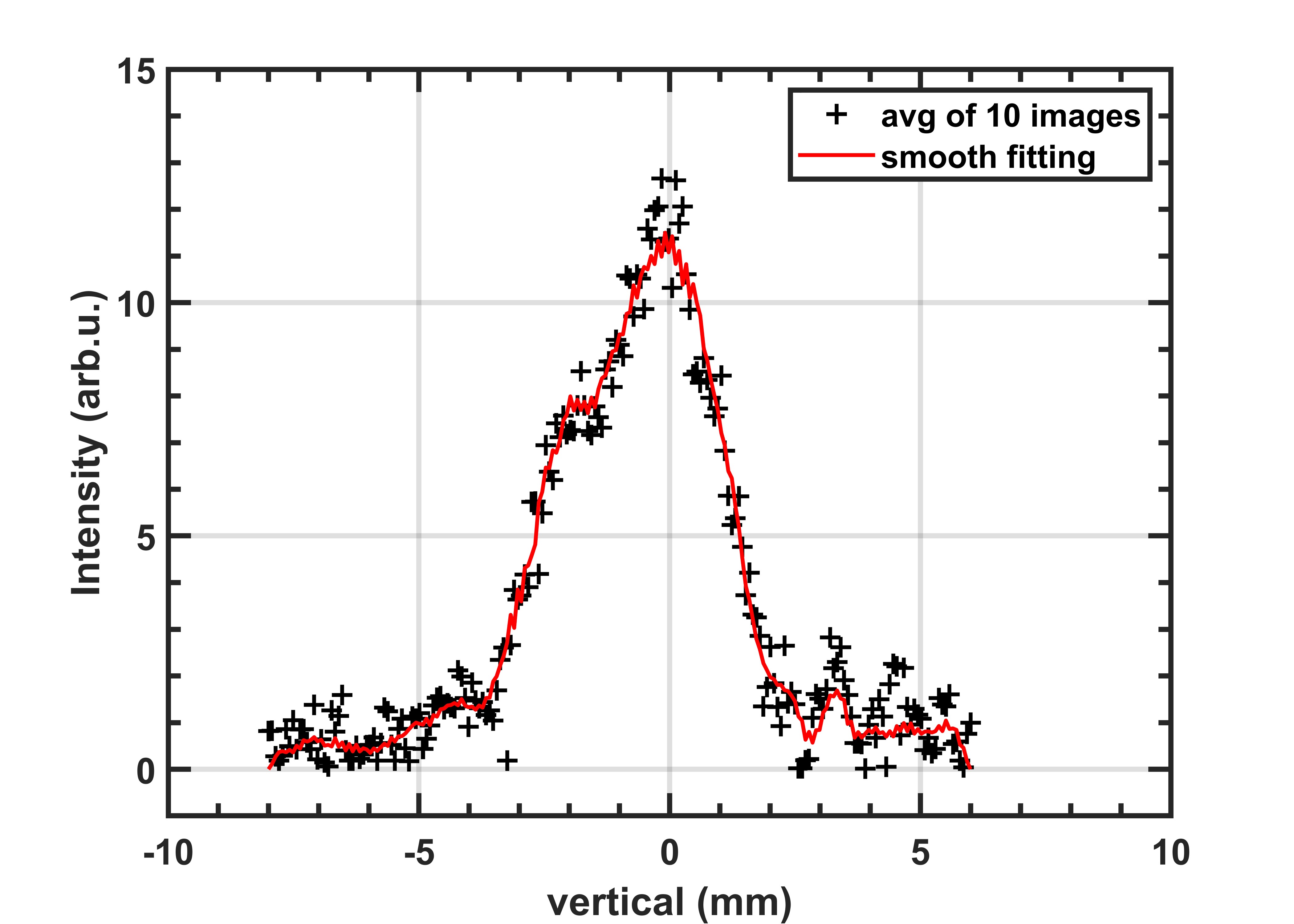}
    \caption{Profile of the 24 MeV C5+ , $\sim$100 nA generated using 10 images of 100 ms integration time each.}
    \label{fig:beam_2D_profile}
\end{figure}

The transverse beam profiles are extracted from the 2 dimentional profile, and the appropriate pixel-to-length scaling is applied to estimate the horizontal and vertical beam sizes. The transverse profiles before and after smoothing are shown together in figure \ref{fig:beam_2D_profile} to demonstrate that smoothing has a minimal effect on the shape. For each beam conditions in Table \ref{tab:beam_current}, profile measurements were repeated for different exposure times (100, 300 and 1000 ms for C$^{5+}$) to validate that the total integrated volume intensity remains linear with the integration time. 

\subsection {Quantifying the sensitivity of IPM}

With emerging modalities like FLASH targeting dose delivery through mini-beams operating in pulse mode\cite{Prezado2013}, online monitoring implies providing pulse-to-pulse information. Hence, the secondary objective was to establish a framework to estimate the sensitivity of the SGC-IPM, allowing us to assess its viability for different beam types. In this work, sensitivity is quantified as the threshold number of ions required to generate a detectable signal by the IPM unit. A quantitative study is carried out in relation to the parameters of the accelerator planned by LhARA collaboration \cite{Aymar2020} which aims to deliver FLASH doses in pulses of a few tens of nanoseconds using proton (average $\sim$15 Gy/sec) and carbon (average $\sim$70 Gy/sec) beams with instantaneous dose rates of the order of $\mathrm{10^9}$ Gy/sec. Online beam monitoring solutions are required for both proton and carbon $\mathrm{C^{6+}}$ beams of energies 15 - 127 MeV, and 33.4 MeV/nucleon, respectively. Here the present experimental results are used to estimate the total number of particles per bunch required for the IPM to detect the individual pulses at the LhARA anticipated beam parameters.

\begin{figure}
    \centering
    \includegraphics[width=0.6\linewidth]{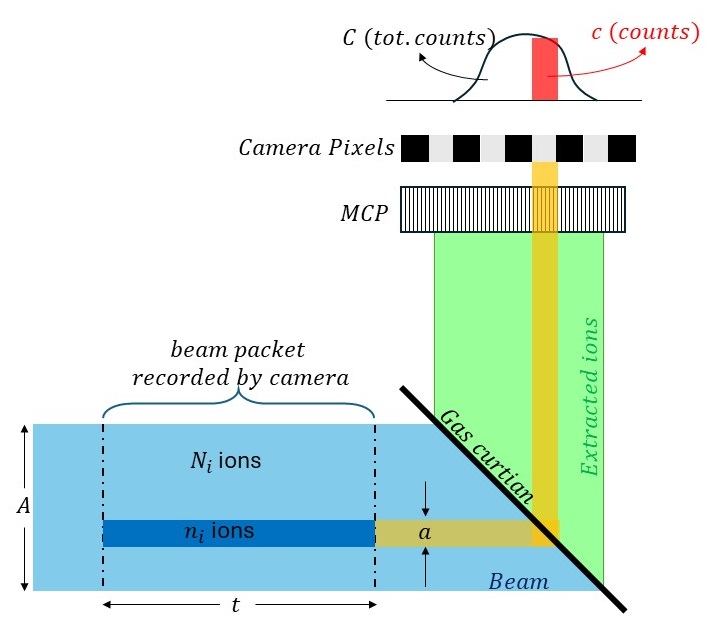}
    \caption{Representation of the detection region of the camera pixel.}
    \label{fig:Dlimit}
\vspace{10mm}
\end{figure}

In the IPM, the uniform electric field of the extraction system ensures that ions originating from ionization events within a fixed region of the gas curtain follow the same trajectory toward the detector. This means a single MCP channel collects ions from a specific region of the curtain, and the circular channel of the MCP subtends a larger elliptical collection area on the gas curtain. Due to the 45$\mathrm{^o}$ inclination, this elliptical area on the curtain translates to a circular area on the beam cross-section, matching the original size of the channel. Thus, only those beam ions crossing the area projected by the MCP channel on the beam cross-section contribute to the signal of that particular MCP channel. We name this area the 'detection area' of the channel. figure \ref{fig:Dlimit} shows the representation of such a detection area. A single pixel of the camera captures multiple MCP channels (approximately 4), and similarly, a detection area can be defined for a single pixel. This area, in our case, is circular with a diameter of 70 µm and shown as 'a' in figure \ref{fig:Dlimit}.

We define the parameter $D$ as the total number of the beam ions per unit detection area of a pixel required to register a single count on that pixel. As shown in figure \ref{fig:beam_profile} (a), at least  5 counts are required to achieve a signal-to-noise ratio greater than one. Hence, this threshold can be defined as $D_{lim}$, which is the number of beam ions per unit cross-sectional area subtended by a single camera pixel on the beam, required to register at least 5 counts on that pixel. This threshold helps estimate the minimum achievable integration time for a given beam current density. While the actual number of ions or integration time needed for a reasonably accurate beam profile is higher than this threshold, knowing this limit informs further enhancements needed to improve the IPM's performance. It should be noted that for a fixed detector gain, $D_{lim}$ depends on the ionization cross-section. Since ionization cross-section depends both on the beam energy and interacting species, $D_{lim}$ values estimated in this work are unique for each combination of beam charge and energy, for a particular density of the nitrogen and argon jets. 

To generate the mathematical model of $D$ for any random beam profile, it is first assumed that the particle flux in the beam follows an unknown distribution function $f(x,y)$ of unit height, where $x$ and $y$ represent the positions relative to the beam center. If the beam profile recorded on the detector is of the same size as the original beam, the subtended area of the pixel will be at the same offset from the beam center as it is on the detector, and the distribution function of the generated profile will follow the distribution function of the beam.  The number of particles $n_i$ crossing the area $a$ substended by a particular pixel $P$ on the beam during the camera's exposure duration $t$ can then be given by Eq. \eqref{eq:n_i}. 

\begin{equation}
    n_i =  \iint_{a} f(x,y) \,dx \,dy \times t 
    \label{eq:n_i}
\end{equation}

The beam profile recorded on the detector can similarly be represented by a function $g(x_p,y_p)$, with $x_p$ and $y_p$ representing the pixel position. The counts $c$ registered on the corresponding pixel $P$ can be represented by Eq. \eqref{eq:c}.

\begin{equation}
    c =  g(x_p,y_p) \times t 
    \label{eq:c}
\end{equation}

As defined earlier, $D$ is the number of beam ions per unit cross-sectional area subtended by a single camera pixel on the beam required to register a single count. Mathematically, this can be defined by Eq. \eqref{eq:D} 

\begin{equation}
    D =  \frac{n_i/a}{c}=\frac{\iint_{a} f(x,y) \,dx \,dy }{g(x_p,y_p) \times a} 
    \label{eq:D}
\end{equation}

This can be simplified by integrating over the entire area $A$ of the beam. The numerator then represents the total number of ions $N_i$  in the beam packet recorded within the exposure time $t$, while the first term in the denominator represents the total integrated counts $C$. Thus, Eq. \eqref{eq:D} simplifies to Eq. \eqref{eq:D_int} as follows.

\begin{equation}
    D =  \frac{ \iint_{A} f(x,y) \,dx\,dy } { \iint_{A} g(x_p,y_p) \,dp \times a} =  \frac{N_i}{C\times a}
    \label{eq:D_int}
\end{equation}

For a given camera exposure time, for continous beams, $N_i$ can be estimated from the known beam current $I$ using Eq. \eqref{eq:n_incident}, where $q$ is the charge state of the ion and $e$ is the elementary charge constant.

\begin{equation}
    N_i=\left( \frac{I}{e \times q} \right) \times t
    \label{eq:n_incident}
\end{equation}

Thus, parameter $D$ can be estimated from the ratio of the total number of particles per beam to the volume under the 2D spatial profile of the beam (figure \ref{fig:beam_profile}(a)), scaled by the area of a single pixel. 

Alternatively, $D$ can be estimated through logical reasoning, as illustrated in figure \ref{fig:Dlimit}. Since a single pixel in the camera subtends a small area on the beam cross-section, the total number of ions $n_i$ contributing to the signal can be calculated by scaling the total ions $N_i$ by the fractional area and multiplying it with a non-uniformity factor $F$ to account for a higher or lesser number of ions than what would be expected for a uniform beam. 

\begin{equation}
    n_i = F \times N_i \times (a/A)
    \label{eq:n_i_simple}
\end{equation}  

The counts per pixel can be similarly estimated from the total integrated counts scaled to the ratio of the measured pixel area to the total area of the 2D beam profile, and then multiplied by the same non-uniformity factor $F$.

\begin{equation}
    c = F \times C \times (1/{Tot.~Pixels~area})
    \label{eq:c_simple}
\end{equation}  

Since the pixel ratio is the same as that of the area ratio, $D$ again equates to $N_i/C \times a$ similar to that in Eq. \eqref{eq:D_int}. It is important to note that $D$ is expressed in terms of the detection area of the pixel, not the beam area itself. This characteristic makes $D$ independent of the beam size; regardless of the beam's spatial extent, the total number of counts registered will be the same under identical conditions. Smaller beams may yield a better signal and be easier to detect, not due to enhanced sensitivity, but simply because all ions contribute to the signal through less pixels. 

For all the beams, the 2D spatial profile similar to the one shown in figure \ref{fig:beam_profile}(a) is generated. The surface is then integrated over the pixel space to estimate the total counts $C$ registered on all the pixels combined. The value of $C$ is also estimated for all combinations of integration time and currents, as well as using multiple images grouped in sets of 1, 10, 20, ... 100. For example, for a single image generated for 100 nA, carbon 5+ beam at 100 ms, the recorded $C$ is approximately $6 \times 10^4$ counts, and the calculated $N_i$ is $1.32 \times 10^{10}$ ions.  Since each pixel subtends a square area of 70 microns, the value of $D$ equates to approximately  $4.4 \times 10^{7}$. The values of $D$ at different beam energies are shown in the Table \ref{tab:D_values}. As expected, the $D$ remains the unchanged for different beam currents and shapes within 3\% uncertainity for the most part. It varies with the charge state and beam energy because of the difference in the ionization cross-section, which affects the the sensitivity. 

\begin{center}
\begin{table}[h]
    \centering
    \caption{$D$ for different beam conditions. Uncertaininy represents the variation accross different beam currents}
    \label{tab:D_values}
    \begin{tabular}{ccccc}
       \hline
	  \textbf{Jet} & \textbf{Beam} & \textbf{D(ions/mm$^{2}$/count)} & \textbf{stddev} & \textbf{\%dev} \\
       Argon 		& 12MeV Carbon 2+ & $1.49\times 10^8$ & $1.68\times 10^6$ & 1.09 \\
	  Argon 	& 16MeV Carbon 3+ & $1.09\times 10^8$ & $8.83\times 10^5$ & 0.81 \\
	  Argon 	& 20MeV Carbon 4+ & $8.58\times 10^7$ & $6.98\times 10^5$ & 0.80 \\
	  Argon 	& 24MeV Carbon 5+ & $4.37\times 10^7$ & $3.15\times 10^5$ & 0.73 \\
	  Nitrogen 	& 16MeV Carbon 3+ & $1.97\times 10^8$ & $7.27\times 10^6$ & 3.21 \\
	  Nitrogen 	& 20MeV Carbon 4+ & $1.69\times 10^8$ & $1.35\times 10^6$ & 0.80 \\
       \hline
    \end{tabular}
\end{table}
\end{center}

\begin{figure}
    \centering
    \includegraphics[width=0.8\linewidth]{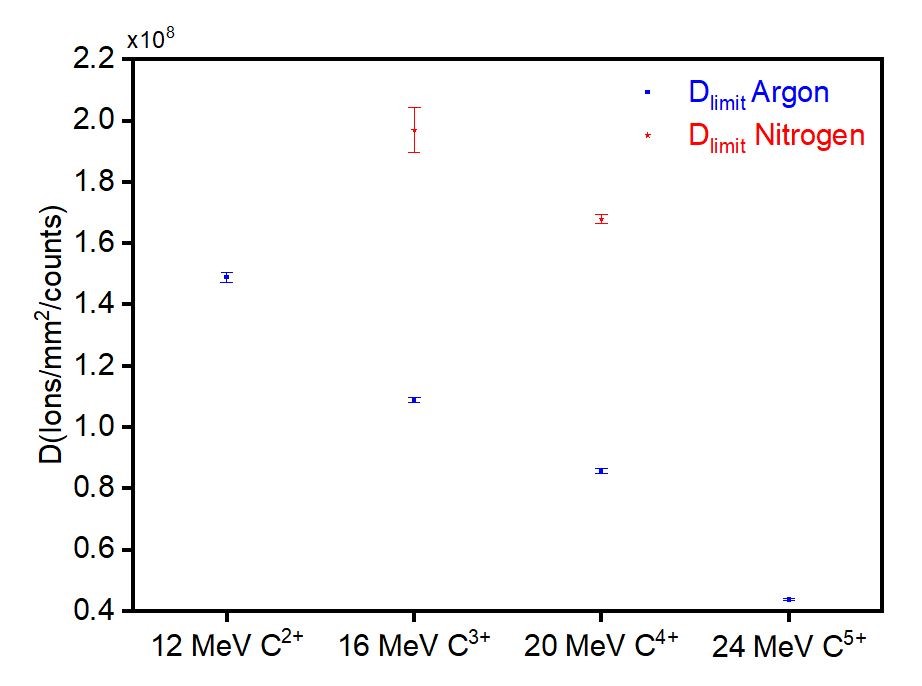}
    \caption{Variation of the minimum detection limit ($\mathrm{D_{limit}}$) with the variation of the charge state and energy of the carbon beam.}
    \label{fig:d_limit}
\vspace{10mm}
\end{figure}

Figure \ref{fig:d_limit} shows the $D$ values for the beam conditions indicated in the Table \ref{tab:D_values}. The trend for $D$ is consistent for both gas species within the available experimental data. The ratio of $D$ for argon relative to nitrogen varies from 0.51 to 0.55. This trend is in statistical agreement with the expected density variation of the gas curtain at the interaction point (0.48), as indicated in figure \ref{fig:gas_jet_density}. An additional factor that could potentially lead to different $D$ values for both gases is the ion impact ionization cross-section. However, the impact ionization cross-section data for various gas targets in the MeV energy range is not available for carbon beams with charged species from 2+ to 5+. Nevertheless, some conclusions can be drawn from the available ionization cross-section data for proton ion beams with energies up to 5 MeV for various gases \cite{Bug2013,Toburen1971,Ziegler1995}. The available data indicates that argon and nitrogen have almost identical ionization cross-section for proton beams in this energy region. Based on the equivalent velocity concept, the ionization cross-section for carbon ion beams with energies from 1-2 MeV/u for argon and nitrogen should be similar to proton beams with energies of 1-2 MeV \cite{Parker2018}. This suggests that the difference in $D$ value is mostly influenced by the density of the gas curtain for the same beam parameters.


\section{Discussions}

\subsection{Comparison with LhARA beam parameters}

The SGC-IPM tested in this work is expected to be installed at the user end-station of the LhARA accelerator, where single pulse detection is required. Hence, it is important to evaluate how the current detection limit would compare with the beam parameters at the LhARA end-station. $D$ quantifies the sensitivity in terms of the number of ions needed for detection, irrespective of their delivery times. Therefore, even though $D$ is evaluated for a continuous beam, it can be compared to pulsed beams provided instantaneous values of beam parameters are considered. The LhARA end-station is expected to deliver approximately $10^8$ ions in a single bunch of 75 ns for a $\mathrm{C^{6+}}$ beam at 33.4 MeV/u. 

The trend of $D$ for 12-24 MeV carbon ions with charge states from +2 to +5 shown in figure \ref{fig:d_limit} can be extrapolated to predict the value of $D$ for a 28 MeV $\mathrm{C^{6+}}$ ion beam (2.33~MeV/u). The predicted $D$ is $9.45 \pm 3.79 \times 10^6$ for argon and $1.85 \pm 0.74 \times 10^7$ for nitrogen. To achieve a signal-to-noise ratio greater than 1, at least 5 counts need to be registered, which defines the limiting value $D_{lim}$ as $D$ multiplied by factor of 5. For argon, $D_{lim}$ would be $4.73 \pm 1.9 \times 10^7$, and for nitrogen, it would be $9.25 \pm 3.7 \times 10^7$. 

The limiting values for 2.33 MeV/u can then be used to infer the limiting values at beam energies of 33.4 MeV/u by correlating the ionization induced in the gas curtain to the stopping power of that gas ($\mathrm{\frac{dE}{dx}}$), which is influenced by gas jet density and the kinetic energy of the incident beam according to the Bethe-Bloch equation. The Bethe-Bloch equation states that the stopping power is inversely proportional to the square of the velocity of the incident beam, indicating a reciprocal relationship with the beam’s kinetic energy \cite{Ziegler1995}. Taking into account the parameters of LhARA, the ionization induced in the gas jet, and consequently the resulting signal, would decrease by approximately a factor of ~14.34 compared to the signal obtained for a 2.33 MeV/u $\mathrm{C^{6+}}$ beam. To achieve a similar signal level, the system would need to enhance the gain by the same factor by which the ionization cross-section would decrease due to the final energy of the beam. The resulting limiting values for a 33.4 MeV/u $\mathrm{C^{6+}}$ beam would then be $6.8 \pm 2.7 \times 10^8$ for argon and $1.3 \pm 0.53 \times 10^9$ for nitrogen. 

A detailed comparision of $D_{limit}$ for the beam parmeters of a pulse $\mathrm{C^{6+}}$beam of different field sizes expected at the user end-station for LhARA \cite{Aymar2020} is summarized in Table \ref{tab:limits}. The table also informs on the additional factor of gain that would be required for the SGC-IPM to measure the transverse beam profile for a single bunch within 100 ms.

\begin{center}
\begin{table} [h]
    \footnotesize
    \centering
    \caption{LhARA Beam Parameters \cite{Aymar2020} and respective performance of the gas jet monitor}
    \label{tab:limits}
    \begin{tabular}{ccc}
        \hline
        \textbf{LhARA Beam parameters} & \multicolumn{2}{c}{$C^{6+} (33 MeV/u)$} \\ 
        \hline
        Instantaneous dose rate (Gy/sec)            & \multicolumn{2}{c}{$9.70 \times 10^8$}    \\
        Pulse length (ns)                   & \multicolumn{2}{c}{75}                    \\
        Repetition rate (Hz)                & \multicolumn{2}{c}{10}                    \\
        Charge per laser shot (pC)          & \multicolumn{2}{c}{100}                   \\
        Ions per bunch (no.)                & \multicolumn{2}{c}{$1.04 \times 10^8$}    \\
        instantaneous beam current (mA)     & \multicolumn{2}{c}{1.33}                  \\
        Square beam of size 3.5×3.5 $cm^{2}$ (ions/sq.mm/bunch)& \multicolumn{2}{c}{$8.49 \times 10^4$}    \\
        Round beam of size 3 $cm^{2}$ (ions/sq.mm/bunch)       & \multicolumn{2}{c}{$1.47 \times 10^5$}    \\
        Pencil beam of size 1 $mm^{2}$ (ions/sq.mm/bunch)      & \multicolumn{2}{c}{$1.32 \times 10^8$}    \\
        \hline
        \textbf{Parameters extrapolated for $C^{6+}$, 2.33 MeV/u}  & Argon jet & Nitrogen jet  \\
        \hline
        $D_{lim}$(ions/sq. mm for 5 counts)      & $4.73 \pm 1.9 \times 10^7$   & $9.25 \pm 3.7 \times 10^7$           \\
        Additional gain required    & \multicolumn{2}{c}{To generate on an average 5 counts/sq.mm}  \\
        for square beam of 3.5×3.5 cm       & 557       & 1090          \\
        for round beam of 3 cm              & 321      & 630           \\
        for pencil beam of 1 mm             & 0.36     & 0.7          \\
        \hline
        \textbf{Parameters calculated for $C^{6+}$, 33.4 MeV/u}  & Argon jet & Nitrogen jet  \\
        \hline
        $D_{lim}$(ions/sq. mm for 5 counts)      & $6.8 \pm 2.7 \times 10^8$   & $1.3 \pm 0.53 \times 10^9$           \\
        Additional gain required    & \multicolumn{2}{c}{To generate on an average 5 counts/sq.mm}  \\
        for square beam of 3.5×3.5 cm       & 8000       & 15312          \\
        for round beam of 3 cm              & 4625       & 8843           \\
        for pencil beam of 1 mm             & 5      & 9.85          \\
        \hline
    \end{tabular}
\end{table}
\end{center}

\subsection{Future enhancements planned}
The present measurement and subsequent data analysis indicate that the current design of the SGC-IPM can measure the transverse beam profile in a single bunch for a pencil beam of $\mathrm{1mm^2}$ at $\mathrm{C^{6+}}$, 2.33 MeV/u energy, and also for 33.4 MeV with a moderate increase in gain. However, to measure larger and less intense beams, significant improvements in gain is required. Major improvements can be achieved by upgrading the existing single plate MCP to a dual plate chevron configuration and replacing the phosphor screen with an anode assembly, which can improve the theoretical gain by a factor of 10$^3$. Increasing the gas curtain density and thickness can increase the gain by a further factor of approximately 5. Some enhancement can also be achieved by selecting gases with higher ionization cross-sections (for example, Xenon). A few of these upgrades will be implemented for the future measurements.

\section{Conclusion}

This paper presents the successful integration and testing of a supersonic gas jet monitor for measuring the transverse beam profile of carbon ion beams under various beam parameters. The results demonstrate that the monitor can measure the beam profile of a ~100 nA carbon beam within ~100 ms. A quantification of the detection limit, based on the current design parameters is proposed, enabling performance comparisons across different beam conditions. The monitor's capabilities were evaluated in the context of the forthcoming LhARA hadron beam facility, which is designed to deliver FLASH beams. Future upgrades aim to further enhance detection capabilities, improve acquisition speed, and extend the device's applicability to a broader range of beam parameters.

\section{Acknowledgement}

This work is supported by STFC Grants ST/W002159/1 and ST/X002632/1, University of Liverpool Faculty Impact Fund, the HL-LHC-UK project funded by STFC and CERN and the STFC Cockcroft core grant No. ST/G008248/1. We acknowledge the support of The University of Manchester’s Dalton Cumbrian Facility (DCF), a partner in the National Nuclear User Facility, the EPSRC UK National Ion Beam Centre and the Henry Royce Institute. We recognize Dr. Andrew Smith, DCF for the assistance during the preparation and execution of the whole experiment.

\clearpage

\section*{References}

\addcontentsline{toc}{section}{\numberline{}References}
\vspace*{-10mm}

\end{document}